\pdfoutput=1
\documentclass{nature}
\usepackage[latin1]{inputenc}
\usepackage{ifthen}
\usepackage[pdftex]{hyperref}
\usepackage[pdftex]{graphicx}
\usepackage{amssymb}
\usepackage{amsmath}

\renewcommand{\u}[1]{\,\mathrm{#1}}

\newcommand{\Rb}{$^{87}$Rb}
\newcommand{\avg}[1]{\langle #1 \rangle}

\newcommand{\Figref}[1]{Fig.~\ref{#1}}

\newcommand{\eprint}[2][e-print]{%
  \ifthenelse{\equal{#1}
             {arXiv}}
             {\href{http://arxiv.org/abs/#2}{\nolinkurl{arXiv:#2}}}{%
  \ifthenelse{\equal{#1}
             {doi}}
             {\href{http://dx.doi.org/#2}{\nolinkurl{doi:#2}}}{%
   \href{#1}{\nolinkurl{:#2}}
}}}

\begin{document}
\title{Spontaneous pattern formation in an anti-ferromagnetic quantum gas}
\author{Jochen Kronj\"ager$^{1,2*}$, Christoph Becker$^1$, Parvis Soltan-Panahi$^1$, Kai Bongs$^2$ and Klaus Sengstock$^1$}
\maketitle

\begin{affiliations}
\item Institut f\"ur Laser-Physik, Universit\"at Hamburg, Luruper Chaussee 149, 22761 Hamburg, Germany
\item MUARC, School of Physics and Astronomy, University of Birmingham,
Edgbaston, Birmingham B15 2TT, UK
\item[$^*$] e-mail: j.kronjaeger@bham.ac.uk
\end{affiliations}

\begin{abstract}
Spontaneous pattern formation is a phenomenon ubiquitous in nature, examples ranging from Rayleigh-Benard convection to the emergence of complex organisms from a single cell. In physical systems, pattern formation is generally associated with the spontaneous breaking of translation symmetry and is closely related to other symmetry-breaking phenomena, of which (anti-)ferromagnetism is a prominent example. Indeed, magnetic pattern formation has been studied extensively in both solid-state materials and classical liquids. Here, we report on the  spontaneous formation of wave-like magnetic patterns in a spinor Bose-Einstein condensate, extending those studies into the domain of quantum gases. We observe characteristic modes across a broad range of the  magnetic field acting as a control parameter. Our measurements link pattern formation in these quantum systems to specific unstable modes obtainable from linear stability analysis. These investigations open new prospects for controlled studies of symmetry breaking and the appearance of structures in the quantum domain. 
\end{abstract}

%
%
Bose-Einstein condensates of alkali atoms such as \Rb\ offer unique opportunities to study classical non-linear effects in the quantum regime \cite{Deng1999a, Campbell2006, Denschlag2000a, Burger1999a, Becker2008}. Governed by many-body quantum mechanics but subject only to weak interactions due to atomic collisions, Bose-Einstein condensates are very well described by a single-particle wavefunction whose dynamics is given by a non-linear Schr\"odinger equation  known as Gross-Pitaevskii equation. Spinor Bose-Einstein condensates \cite{Ho1998a,Ohmi1998a} in addition make use of the non-zero spin and associated magnetic moment of alkali atoms in their internal ground state. Adding the orientation of this spin as a degree of freedom, non-trivial dynamics arises from the spin-dependence of collisional interactions on the one hand, and interaction of the atomic moments with an external magnetic field on the other hand. Although based on fundamentally different interactions this mechanism effectively adds magnetic properties to the quantum gas, similar to the ones in solid state systems. 

%
%
Indeed the dominant spin dependent interaction term in ultracold atomic collisions is proportional to the product of hyperfine spins $\vec{F_1} \cdot \vec{F_2}$ and thus reminiscent of the Heisenberg model of solid-state magnetism -- this analogy motivated the classification of spinor BEC ground states  as \emph{ferromagnetic} and \emph{anti-ferromagnetic} in previous experiments \cite{Stenger1998a,Chang2004a,Schmaljohann2004b}. However, it is worth noting  that interactions of ultracold atoms are purely local (usually approximated by a $\delta$-potential), in contrast to the next-neighbor interactions common in solid-state models. Further differences  arise from the hyperfine nature of the atomic spin, i.e. its composition of nuclear and electronic spins. First of all, its value can be significantly larger than the electron spin $S=1/2$, implying a higher dimensional state space with potentially much richer structure and dynamics. Secondly, the internal structure of hyperfine states leads to a more complex interaction with external magnetic fields, beyond the linear Zeeman effect.  The competition between spin-dependent collisional interactions and quadratic Zeeman effect lies at the heart of spinor BEC physics. It is the driving force behind coherent oscillations in quasi-homogeneous systems \cite{Chang2005a,Kronjaeger2006}, and it is responsible for the quantum phase transition associated with the formation of irregular spatial spin structures \cite{Zhang2005b,Sadler2006a}.

%
%
The wavefunction of a spin-$F$ condensate can be written in terms of a density and a local spinor, $\psi_m(x) = \sqrt{n(x)} \zeta_m(x)$, normalized to the total number of particles $\int n(x)\,dx = N$ and unity $\sum_{m=-F}^F |\zeta_m|^2 = 1$ respectively. In the simplest case of a spin-1 condensate, the local mean-field energy associated with $\psi$ is given by $g_0n + g_1n\avg{\vec{F}}^2$. \footnote{Spin-2 condensates have an additional contribution $g_2n|S_0|^2$ where $S_0 = \zeta_{+2}\zeta_{-2} - \zeta_{+1}\zeta_{-1} + \zeta_0^2/2$. This contribution is included in our numerical calculations in the low-field case, however its influence is minor and we neglect it in this Letter for the sake of simplicity.} The coefficients $g_{0,1}$ are determined by the atomic s-wave scattering lengths. The $g_1$ term represents the spin-dependent part of the interactions, $\avg{\vec{F}}$ is the local expectation value of the spin vector. This term generally is a small fraction of the total mean-field energy. For \Rb\ in the $F=2$ state, $g_1$ is positive, indicating an \emph{anti-ferromagnetic} ground state with $\avg{\vec{F}}=0$. In contrast, for $F=1$ $g_1$ is negative, favoring a maximally stretched state (ferromagnetic ground state). Further spin-dependent contributions arise from the interaction of the atomic momenta with an external magnetic field, $- p\avg{F_z} + q\avg{F_z^2}$. The linear and quadratic Zeeman effect are parametrized by $p\propto B$ and $q\propto B^2$ respectively. In the absence of inhomogeneous fields, the linear Zeeman effect merely causes Larmor precession of the spin which disappears in a suitable co-rotating coordinate system. Consequently, all non-trivial dynamics of spinor condensates is driven be the competition of interaction energy and quadratic Zeeman effect.

%
%
In our experiment, we prepare \Rb\ $F=2$ atoms in the fully magnetized state in a \emph{transverse direction}, such that the spin vector is $\avg{\vec{F}} = 2 \vec{e}_x$ in the co-rotating frame (see \Figref{Sketch} (a)). Neglecting the spatial degree of freedom, this state \emph{maximizes} the mean-field energy and is therefore stationary at zero magnetic field, just as the anti-ferromagnetic ground state which minimizes the mean-field energy. However, analogous to an inverted pendulum, the fully magnetized state is stationary but energetically unstable, and thus an ideal starting point for dynamical instabilities and pattern formation. 

%
%
In order to  study \emph{spatial} pattern formation in the simplest non-trivial case, we have developed a reliable method to create extremely elongated Bose-Einstein condensates in the optical potential  of a single focused laser beam, with an axial trap frequency of only $\omega_z = 2\pi\times 0.8\u{Hz}$. The condensates are approximately $800\u{\mu m}$ long and the calculated Thomas-Fermi radii perpendicular to the long axis are $2.9\u{\mu m}$ horizontally and $4.5\u{\mu m}$ vertically. In the transverse direction, the spin healing length $\xi_s = \sqrt{\frac{\hbar}{2mg_1\avg{n}}}\approx 3.4\u{\mu m}$ is of the order of the size of the condensate, effectively suppressing transverse spin structure and creating a quasi-1D geometry.  A sufficiently homogeneous magnetic field, whose axial gradient is canceled with a remaining curvature of $60\u{mG\,cm^{-2}}$, is aligned with the axis of the trap. Stern-Gerlach separation during a short time of flight allows us to image the individual spin densities $n|\zeta_m|^2$ with a spatial resolution of about $3\u{\mu m}$. Image distortions caused by imperfect Stern-Gerlach fields are reversed by careful post-processing.

\begin{figure}
\includegraphics{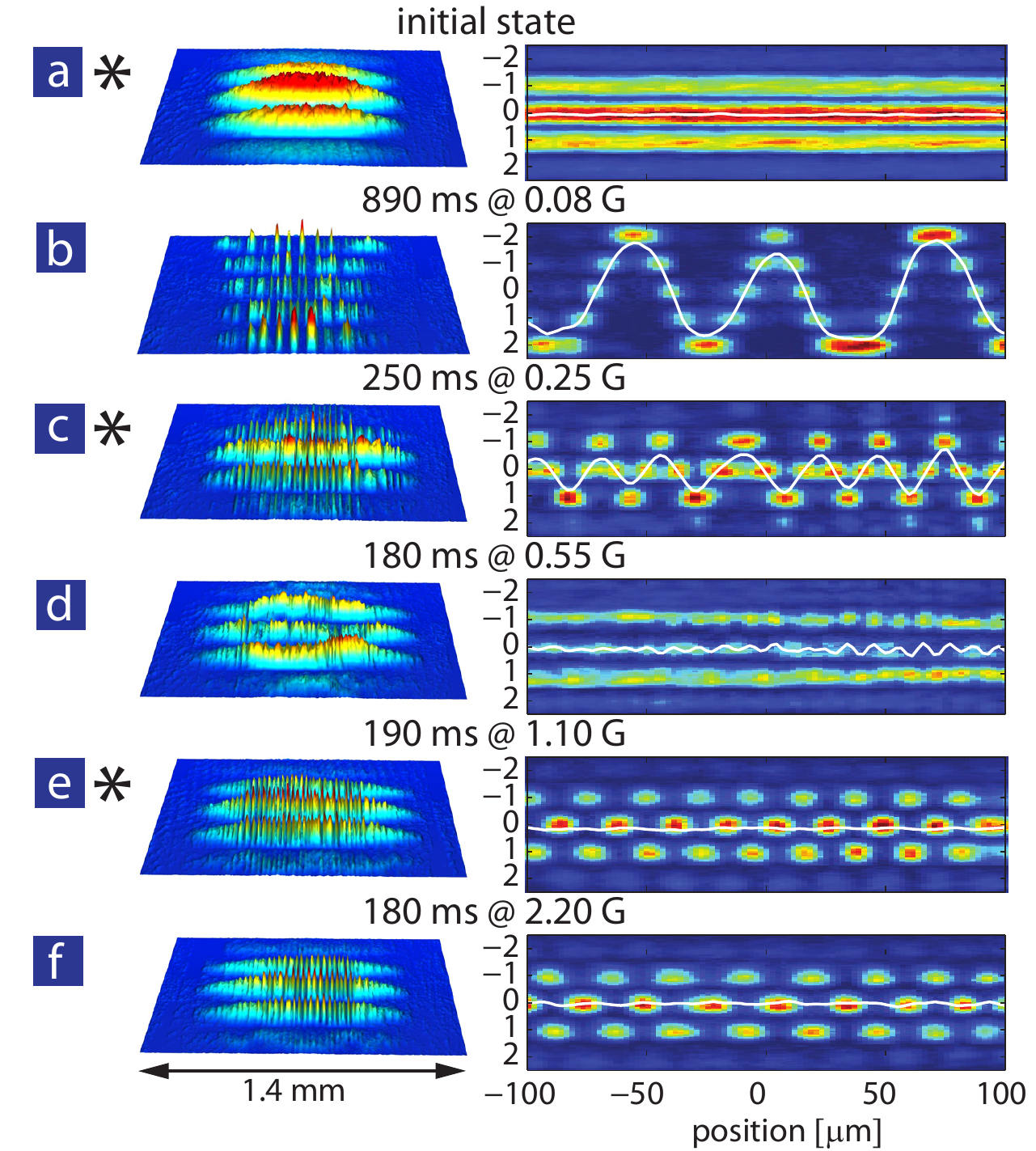}
\caption{\label{Patterns}
Initial state (a) and saturated spin patterns (see text) arising in spinor condensates for increasing magnetic field (b-f). The individual 3D rendered false color absorption profiles in each image visualize the spatial density distribution for each of the five magnetic projections $m_F=-2\ldots+2$ of the $F=2$ hyperfine manifold. Qualitatively different wave-like patterns arise both in the regimes of low and high magnetic field, which are separated by an irregular intermediate regime (d). Imperfections in the preparation and detection artefacts (e.g. optical interference fringes), which appear as fluctuations visible in the initial state (a), are distinctly different from the patterns of (b-f). The cases marked `*' are representative of the initial state, the interaction dominated regime and the Zeeman dominated regime, respectively, and are referred to in the following figures.}
\end{figure}

\begin{figure}
\includegraphics{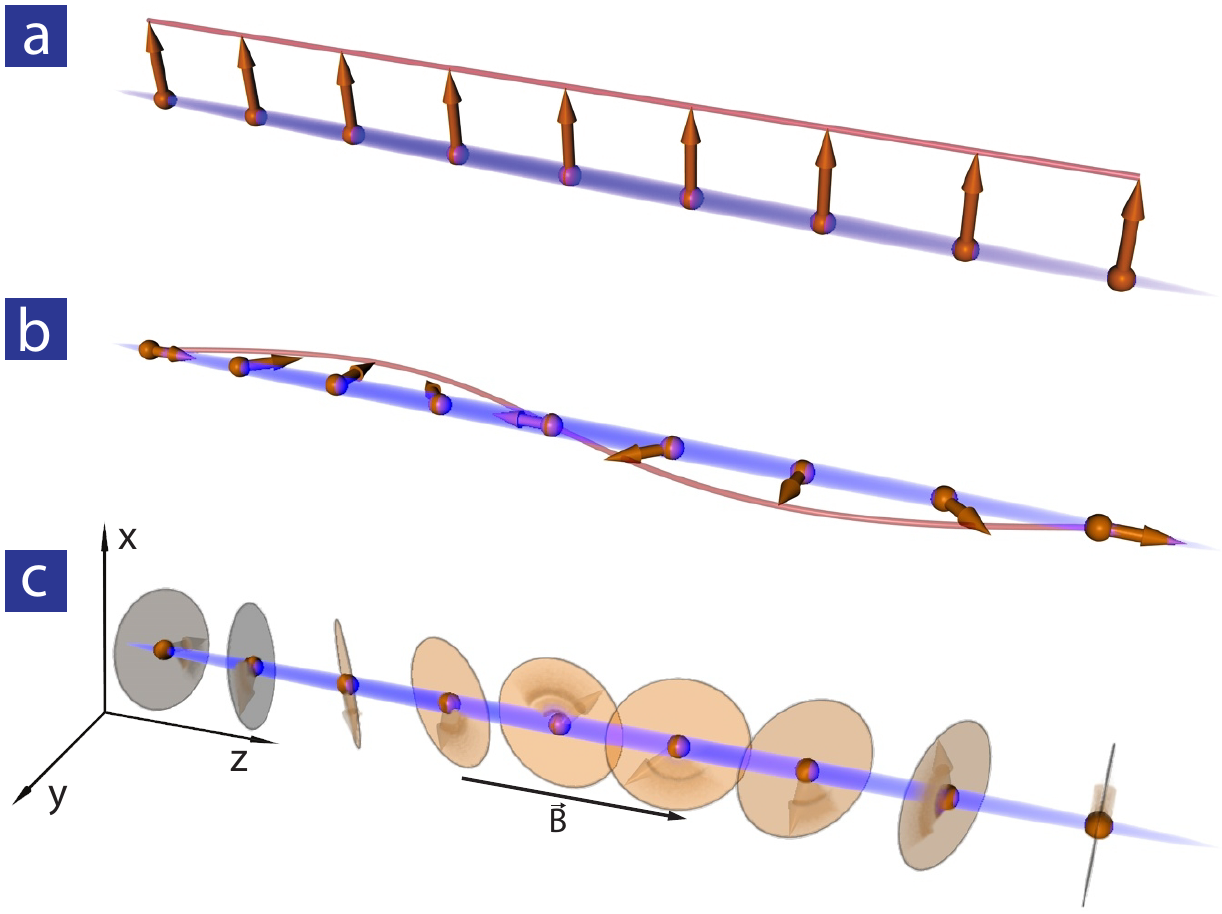}
\caption{\label{Sketch}
Simplified illustration of the local spin vector in the \Figref{Patterns} (a), (c) and (e) (marked `*'). 
(a) Initial state, the spin vector is fully stretched and points in the transverse $x$ direction. 
(b) Interaction dominated regime, the spin vector is fully stretched but rotates from axial to transverse and back over one spatial period. 
(c) Zeeman dominated regime, the local spin vector has zero length but still a defined orientation ($m_F=0$ state) that rotates from axial to transverse and back.
Compare also to the more realistic spin representation in \Figref{Blochsphere}.}
\end{figure}

%
%
We observe the spontaneous emergence of wave-like  spin textures along the axis across a wide range of magnetic fields. \Figref{Patterns} provides an impression of these patterns as they emerge, \Figref{Sketch} shows a possible interpretation in terms of local spin orientation. One can clearly identify spin textures with well defined symmetry and wavelength for both low and high magnetic fields, while the structure is more complicated in an  intermediate regime. At low magnetic field, where the interaction energy dominates ($q \ll g_1 n$),  the mode is anti-symmetric with respect to positive and negative $m_F$, indicating an alternating non-zero axial magnetization. Our data seems to suggest that both the wavelength and growth rate scale with the magnetic field. \footnote{Note that the slight residual axial magnetic field variation in our experiment could possibly influence the patterns observed at the lowest magnetic field $B = 80\u{mG}$.} At high magnetic fields where the quadratic Zeeman energy far outweighs the interaction energy scale ($q \gg g_1n$), mainly the $m_F=0,\pm 1$ spin components are modulated and the characteristic mode is symmetric with respect to $m_F = +1$ and $m_F=-1$.  The axial component of the spin vector $F_z$ thus remains zero everywhere. The wavelength as well as the timescale of emergence of these patterns in the Zeeman regime do not change significantly as the quadratic Zeeman energy changes by a factor of four.

\begin{figure}
\includegraphics{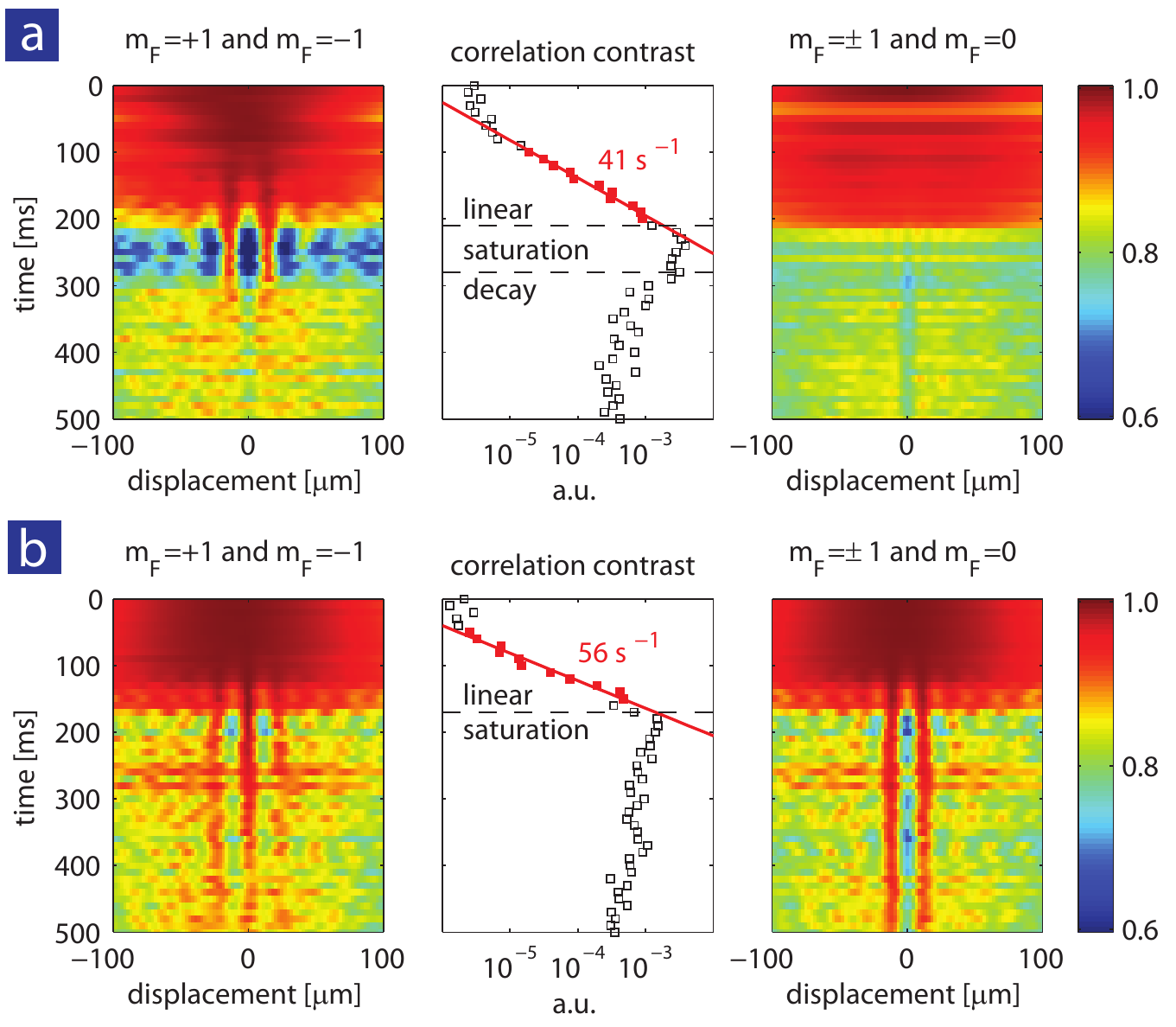}
\caption{\label{Correlations}
Cross-correlations of spin densities. (a) Interaction dominated regime $0.25\u{G}$, (b) Zeeman dominated regime $1.1\u{G}$. 
Color-coded plots (left and right) of the normalized correlation as a function of relative axial displacement and time of evolution  allow to determine the symmetry and  spatial periodicity of the evolving modes. In (a), the mode is  antisymmetric with respect to exchange of $m_F=\pm 1$ whereas the $m_F=0$ component is not involved in the pattern formation. In (b), the mode is symmetric  with respect to $m_F\pm 1$ and antisymmetric with respect to $m_F = 0$ and both $m_F=\pm 1$. 
The correlation contrast (center) can be tracked over two orders of magnitude and allows us to extract the growth rate of the mode and the identification of linear, saturation and decay regimes.}
\end{figure}

%
%
Calculating cross-correlation functions of the individual spin densities allows to evaluate quantitatively the periodicity of the modes as well as the dynamics of their growth (\Figref{Correlations}). At $B=0.25\u{G}$ and $B=1.1\u{G}$, representative of the regimes $q\ll g_1n$ and $q \gg g_1n$ respectively, the wavelength is of the order of $2\pi\,\times$ the spin healing length $\xi_s=3.4\u{\mu m}$ in our experiment, namely $28\u{\mu m}$ and $22\u{\mu m}$ respectively. The growth rate of the mode amplitude is  half the measured growth rate of the correlation function, i.e. $20\u{s^{-1}}$ and $28\u{s^{-1}}$ respectively. In both cases the patterns saturate after about $200\u{ms}$. In contrast to classic pattern-forming system, the patterns developing in elongated spinor condensates are transient and decay into chaotic small-scale structure on the time scale of several hundred milliseconds. Whether this is an intrinsic effect driven by the same nonlinear coupling that leads to the emergence of regular structure in the first place is an unresolved question. Other reasons for the observed decay may be the limited lifetime of atoms in the trap, leading to a loss of density, or additional mechanisms of interaction, such a dipolar forces \cite{Vengalattore2008a}.

\begin{figure}
\includegraphics{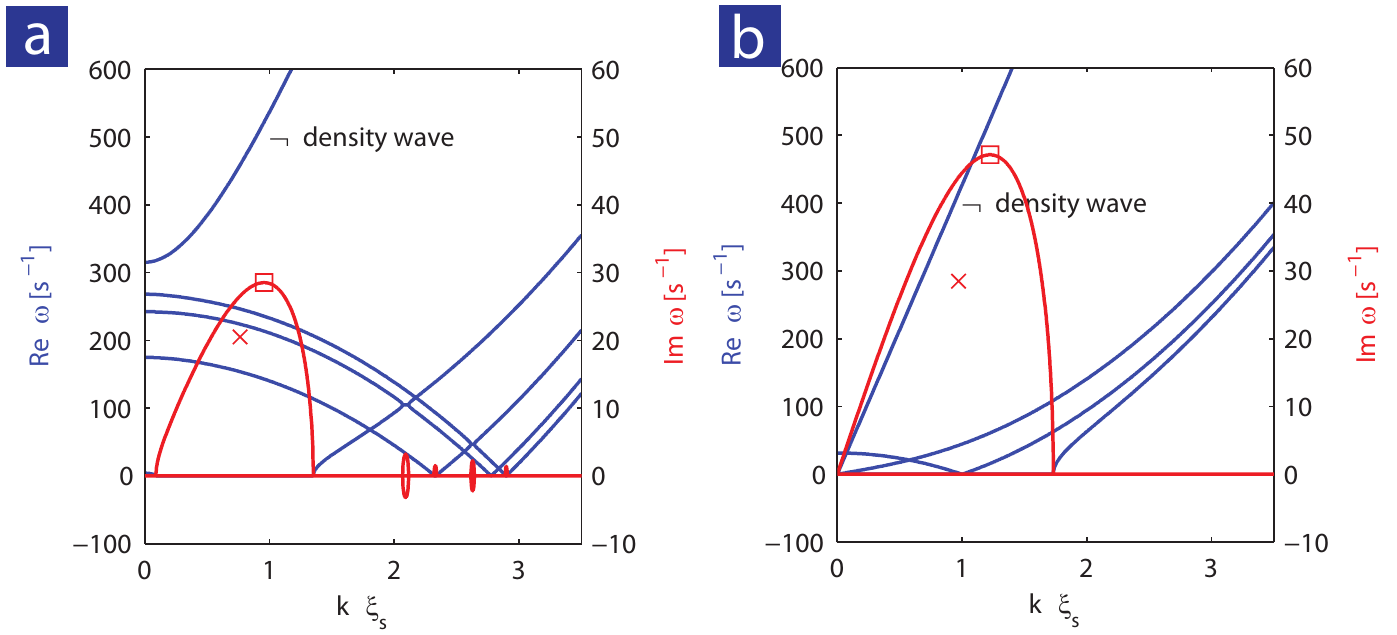}
\caption{\label{Spectra}
Bogoliubov spectra resulting from a linear stability analysis of a homogeneous system, (a) Interaction dominated regime $0.25\u{G}$, (b) Zeeman dominated regime $\infty\u{G}$. For $F=2$, five pairs of complex conjugate branches exist (of which only the positive real frequencies are plotted). In both cases  the highest real frequency modes correspond to density waves (no spin modulation). Branches with positive imaginary frequency indicate unstable modes. The experimentally observed modes (crosses) are qualitatively reproduced, but occur at slightly lower wave vector and have a lower growth rate than the most unstable modes (squares), which we attributed to the approximations in the model.}
\end{figure}

\begin{figure}
\includegraphics{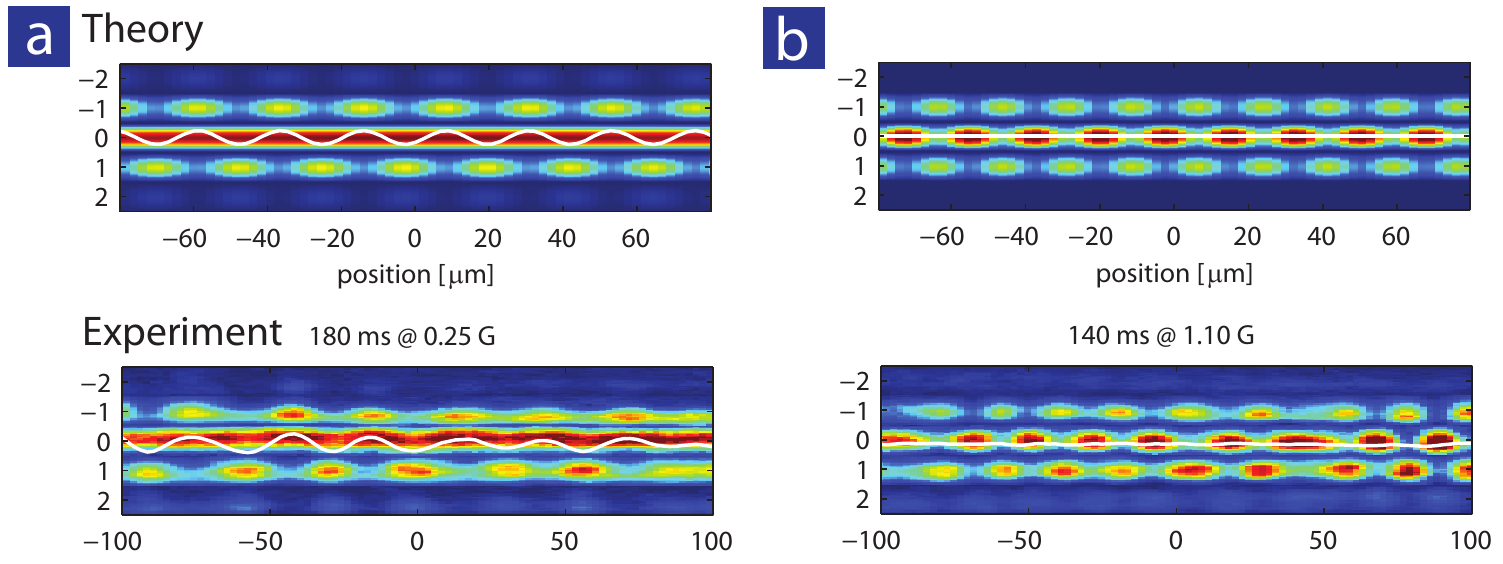}
\caption{\label{SimulatedSG}
Comparison of  simulated Stern-Gerlach images with experimental data in the linear growth regime, (a) interaction dominated regime $0.25\u{G}$, (b) Zeeman dominated regime $\infty\u{G}$. This figure demonstrates that the mode pattern of the calculated most unstable mode actually conincides with with the experimentally observed ones during linear growth, despite quantitative deviations in wavelength and growth rate.}
\end{figure}

%
%
We compare our observations to the Bogoliubov excitation spectrum of a homogeneous, one-dimensional spinor condensate prepared  near the transversely magnetized state (\Figref{Spectra}). It turns out that both in the interaction- and Zeeman-dominated regime singular unstable modes exist, indicated by an imaginary frequency over a certain range of wavevectors. The characteristic spin patterns of these modes indeed coincide very well with the patterns observed experimentally (\Figref{SimulatedSG}). Quantitatively, the calculated wavelengths are 20\% shorter than observed and the growth rates 40-65\% higher. In light of the heavy approximations made in the calculation and the inherently limited validity of the Bogoliubov linearization, we consider this a reasonable agreement.

\begin{figure}
\includegraphics{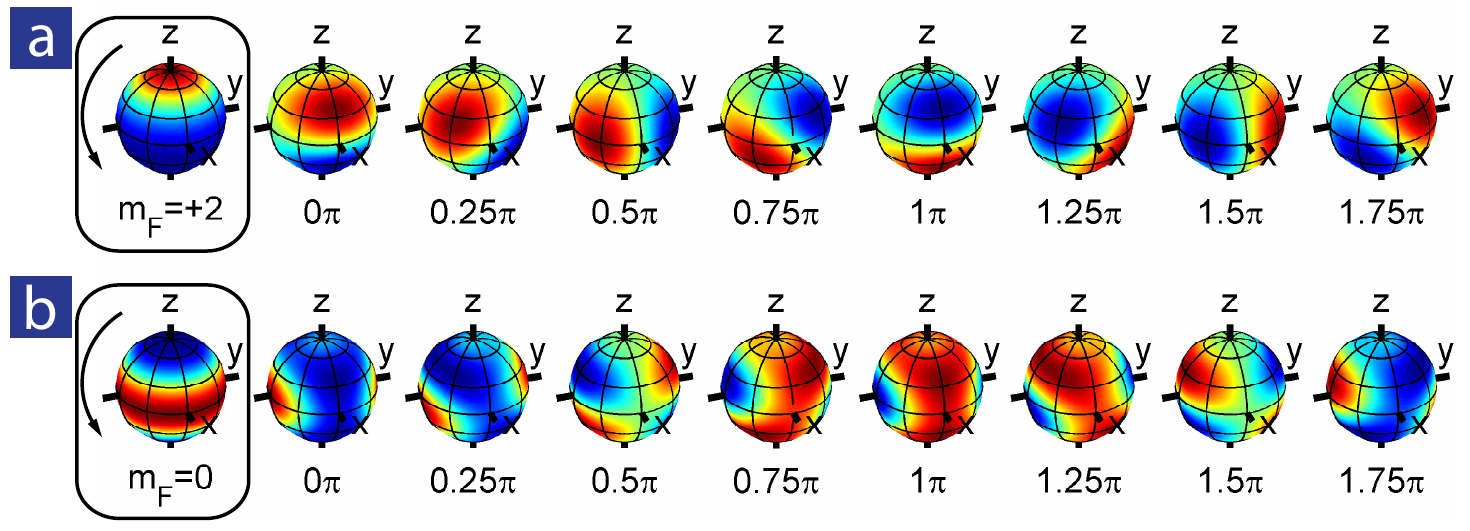}
\caption{\label{Blochsphere} 
Bloch sphere projections of the calculated unstable mode, scaled to span the full color range from blue (minimum) to red (maximum). (a) Interaction dominated regime $0.25\u{G}$, (b) Zeeman dominated regime $\infty\u{G}$. For comparison, the Bloch sphere projections corresponding to \Figref{Sketch} are also displayed (boxes on the left-hand side). These projections confirm the general features of the spin vector illustration in \Figref{Sketch} but also show some differences. In the interaction dominated regime, compared to a fully stretched state rotating around the $x$ axis, the actual mode rotates on a cone rather in the $yz$ plane. The fully axially magnetized state is thus not reached in the linear regime. In the Zeeman dominated regime, the mode does not have the full $m_F=0$ symmetry and shows some additional modulations, however it is still characterized by a similar plane of symmetry. }
\end{figure}

%
%
\Figref{Blochsphere} again illustrates the maximally unstable Bogoliubov modes by plotting their overlap with the Bloch sphere of all stretched states \cite{Barnett2006a}. In the interaction-dominated regime, the mode can be qualitatively characterized (as already sketched in \Figref{Sketch} (b)) by a spin vector $\vec{\avg{F}}$ rotating around the $y$-axis, with a full rotation corresponding to one spatial period of the mode. Thus, the axial and transverse magnetization oscillate out of phase to each other, on top of the fixed transverse magnetization of the initially prepared state. In the Zeeman-dominated regime, the evolution of the spin along the condensate axis is more complex and the local spin state generally lacks the rotational symmetry implied in \Figref{Sketch} (c). However it is still characterized by a plane of symmetry whose orientation rotates as the wave propagates. 

%
%
While linear stability analysis provides a satisfactory explanation of the variety and dynamics of modes observed in experiment, it is a purely classical approach in character. But our work also is a first step towards fundamental quantum mechanical aspects of ultracold matter. For example, having observed exponential growth, the question immediately arises what is the origin of the fluctuations triggering this growth \cite{Klempt2009a, Leslie2008a}. A back-of-the-envelope calculation assuming 100\% modulation after $200\u{ms}$ and a growth rate of $20\u{s^{-1}}$ provides an estimated modulation depth of 2\% initially. At our linear density of about $1.7\times 10^7\u{cm^{-1}}$, this corresponds to four times the shot-noise of the number of atoms contained within one wavelength of $28\u{\mu m}$. Another interesting  issue arises from the similarities between coherent dynamics in spinor BEC and optical non-linearities. Interpreting density patterns as interference between left- and right-propagating modes suggests the possibility of entanglement as in spontaneous parametric down-conversion. Finally, our results also demonstrate that purely local anti-ferromagnetic interactions suffice to generate ordered spin textures, in contrast to recent observations in ferromagnetic \Rb\ $F=1$, where these textures were attributed to dipolar interactions \cite{Vengalattore2009a}.

\begin{methods}
\subsection{Experimental sequence}
Pre-cooled \Rb\ atoms in the $F=1, m_F=-1$ state  are loaded from a magnetic trap into the single-beam dipole trap and evaporatively cooled to quantum degeneracy by ramping down the optical power over $13\u{s}$. During the dipole trap ramp, the magnetic field is set to its final value. The atoms are then prepared in the $F=2, m_F=-2$ state by a microwave pulse and subsequently rotated into the fully transversely magnetized state by a radio-frequency pulse. After a variable time of evolution, the trap is switched off, the magnetic field is non-adiabatically switched to a high value and the Stern-Gerlach gradient field is applied. The total time of flight before absorption imaging is $12\u{ms}$. 
\subsection{Image post-processing}
Linear and non-linear transformations are applied to the absorption images to remove distortions introduced by the Stern-Gerlach procedure. The parameters of the transformations are independently determined from images obtained in the same way, but with regular patterns induced by a defined magnetic field gradient. Axial density profiles of individual $m_F$ components are obtained by transverse integration of the absorption images. Correlation functions are then calculated in two different ways. The color-coded plots in \Figref{Correlations} show normalized correlation functions of density profiles, i.e. at $t=0$ the overall density profile of the condensate can be seen, and at $\Delta z = 0$ auto-correlations equal unity by definition. This method has proven useful for visually bringing out the emerging patterns. Growth rates in contrast are extracted from relative spin populations, in order to avoid the influence of the overall density profile. The correlation functions are not normalized to avoid artefacts. After spatial bandpass filtering centered around the observed periodicity, an rms amplitude is extracted and serves as a measure of the modulation contrast of the pattern.
\subsection{Calibration of parameters}
Transverse residual magnetic fields are carefully compensated to a level of $0.5\u{mG}$ or less, while the axial magnetic field $B_z$ is accurately measured using rf spectroscopy. The axial gradient of $B_z$ is zeroed at the center of the atomic cloud, with a remaining variation $\Delta B_z \approx 35\u{\mu G}$ due to curvature at the ends of the cloud. Over the central $200\u{\mu m}$ region displayed in \Figref{Patterns}, the field-induced phase variation is less than $\pi$ for more than $200\u{ms}$. The average density of condensate atoms in the central region $\avg{n} = 6.5\times 10^{13}\u{cm^{-3}}$ is obtained by comparing the global spin dynamics before the onset of pattern formation with the results of \cite{Kronjaeger2006}.
\subsection{Linear stability analysis}
For low magnetic fields, we analytically linearize the homogeneous, one-dimensional, $F=2$ Gross-Pitaevski equations using a Bogoliubov ansatz and assuming a stationary initial state. For a given set of parameters ($g_1, g_2, q)$, a stationary state close to the fully transversely magnetized state is obtained numerically from the full non-linear equations of motion, and the corresponding Bogoliubov matrices are numerically diagonalized to obtain the spectrum. For high magnetic fields, we eliminate the fast dynamics induced by the quadratic Zeeman effect by applying a unitary transformation $U = \exp(-i q  F_z^2 t / \hbar)$ and dropping all explicitly time dependend terms from the linearized equations. This approximation fully removes the magnetic field dependence and allows us to analytically find another stationary state close to the the fully transversely magnetized state. 
\end{methods}

\begin{addendum}
\item We thank M. Baumert (University of Birmingham) for the 3D artwork in \Figref{Patterns} and \Figref{Sketch} as well as help with graphic design in general. We thank the Deutsche  Forschungsgemeinschaft for support within the Forschergruppe FOR801. P.S. acknowledges support through DFG GRK 1355. K.B. thanks the EPSRC for financial support throught grant EP/E036473/1.
\end{addendum}

\newpage

\end{document}